# Experimental signatures of emergent quantum electrodynamics in $Pr_2Hf_2O_7$

Romain Sibille[1,2], Nicolas Gauthier[2], Han Yan[3], Monica Ciomaga Hatnean[4], Jacques Ollivier[5], Barry Winn[6], Uwe Filges[2], Geetha Balakrishnan[4], Michel Kenzelmann[2,1], Nic Shannon[3] & Tom Fennell[1]

[1]Laboratory for Neutron Scattering and Imaging, Paul Scherrer Institut, 5232 Villigen PSI, Switzerland, [2]Laboratory for Scientific Developments and Novel Materials, Paul Scherrer Institut, 5232 Villigen PSI, Switzerland, [3]Okinawa Institute of Science and Technology Graduate University, Onna-son, Okinawa 904-0495, Japan, [4]Physics Department, University of Warwick, Coventry, CV4 7AL, UK, [5]Institut Laue-Langevin, CS 20156, F-38042 Grenoble Cedex 9, France, [6]Quantum Condensed Matter Division, Oak Ridge National Laboratory, Oak Ridge, Tennessee, USA, *email: romain.sibille@psi.ch

**In a quantum spin liquid, the magnetic moments of the constituent electron spins evade classical long-range order to form an exotic state that is quantum entangled and coherent over macroscopic length scales[1-2]. Such phases offer promising perspectives for device applications in quantum information technologies, and their study can reveal fundamentally novel physics in quantum matter.** *Quantum spin ice* **is an appealing proposal of one such state, in which the fundamental ground state properties and excitations are described by an emergent** *U*(1) **lattice gauge theory[3-7]. This quantum-coherent regime has quasiparticles that are predicted to behave like magnetic and electric monopoles, along with a gauge boson playing the role of an artificial photon. However, this emergent lattice quantum electrodynamics has proved elusive in experiments. Here we report neutron scattering measurements of the rare-earth pyrochlore magnet $Pr_2Hf_2O_7$ that provide evidence for a** *quantum spin ice* **ground state. We find a quasi-elastic structure factor with pinch points – a signature of a classical spin ice – that are partially suppressed, as expected in the quantum-coherent regime of the lattice field theory at finite temperature. Our result allows an estimate for the speed of light associated with magnetic photon excitations. We also reveal a continuum of inelastic spin excitations, which resemble predictions for the fractionalized, topological excitations of a quantum spin ice. Taken together, these two signatures suggest that the low-energy physics of $Pr_2Hf_2O_7$ can be described by emergent quantum electrodynamics. If confirmed, the observation of a quantum spin ice ground state would constitute a concrete example of a three-dimensional quantum spin liquid – a topical state of matter which has so far mostly been explored in lower dimensionalities.**

The idea of a spin system that lacks symmetry-breaking magnetic order at zero temperature traces back to Anderson's 1973 proposal in which valence bonds between neighbouring spins pair into singlets and resonate on the lattice[8]. The idea was then extended to valence bonds at all length scales, leading to a macroscopically quantum entangled ground state wave-function and a quantum spin liquid (QSL)[1-2]. Models supporting highly-entangled QSLs have been developed for various low-dimensional and/or frustrated model systems[1-2,9-10]. An exciting aspect is the emergence of exotic excitations carried by such phases, which behave as quasiparticles and can only be produced by an infinite product of local spin operators – a direct consequence of the many-body entanglement. For instance, the excitations of antiferromagnetic spin-half ($S = 1/2$) chains are deconfined spinons, each carrying $S = 1/2$ – fractionalized quasiparticles that are fundamentally different to the $S = 1$ magnons of conventional three-dimensionally ordered magnets. They have been measured as continua of spin excitations in neutron scattering experiments on one-dimensional magnets such as $KCuF_3$[11]. The physics of fractionalization is also visible in two-dimensional magnets, e.g. kagome-lattice $ZnCu_3(OD)_2Cl_2$[12], honeycomb-lattice α-$RuCl_3$[13] and triangular-lattice $YbMgGaO_4$[14]. However, the experimental evidence for a QSL state stabilized in three dimensions is still absent and the signatures of QSL states that allow direct comparison with theoretical models remain rather elusive.

In three dimensions, spin ice[15,16] is a well-established paradigm to stabilize *classical* spin liquids. In rare-earth pyrochlore materials such as $Ho_2Ti_2O_7$, Ising-like magnetic moments decorate a lattice of corner-sharing tetrahedra. A local constraint – the 2-in-2-out 'ice rule' acting on each tetrahedron – leads to a manifold of degenerate ground states where the spin correlations create an emergent gauge field, $\mathcal{A}(\boldsymbol{r})$ in real space. The ice rule endows the gauge field with a zero-divergence condition that can be written $\mathcal{B}(\boldsymbol{r}) = \nabla \times \mathcal{A}(r)$, where $\mathcal{B}(\boldsymbol{r})$ has the physical meaning of the local magnetization. The result is a Coulomb phase[17,18] where spin flips violating the ice rule generate magnetic monopoles[19], a mobile magnetic charge regarded as a quasiparticle carrying half of the dipole moment, which interact by emergent classical magnetostatics. Specific signatures of the emergent gauge symmetry appear in neutron scattering experiments on spin ices, in which the diffuse response typical of spin liquids acquires pinch points at specific wave-vectors[17,18]. Currently, the quantum spin ice (QSI) state forms a formidable challenge[6]. This type of QSL is a generalization of the classical spin ice that includes quantum fluctuations, whose effective field theory becomes emergent quantum electrodynamics[3,4]. Time fluctuations of the gauge field $\mathcal{A}(\boldsymbol{r})$ give rise to an electric field, $\mathcal{E}(\boldsymbol{r})$, and the ground state is governed by the Maxwell equation

$$\mathcal{S}_{Maxwell} = \frac{1}{8\pi} \int dt d^3\boldsymbol{r} [\mathcal{E}(\boldsymbol{r})^2 - c^2 \mathcal{B}(\boldsymbol{r})^2],$$

which supports linearly dispersing transverse excitations of the gauge field, i.e. emergent photons with a speed of light $c$. The gauge theory of the QSI also supports gapped excitations, which are magnetic monopoles – akin to spinons – and electric monopoles, all described as quantum coherent quasiparticles.

Formally, the QSI can be constructed by introducing transverse terms into an effective $S = 1/2$ Hamiltonian on the pyrochlore lattice[20,21], so that fluctuations between the degenerate ice rule states become allowed via quantum tunneling[3-6]. It was proposed that, in praseodymium-based pyrochlore materials, multipolar couplings can introduce effective transverse exchange couplings between the low-energy effective $S = 1/2$ moments[22]. Experimental investigations on $Pr_2Zr_2O_7$ conform to this theoretical proposal[23], but the results appear affected by structural disorder that brings additional effects into play[24,25]. It is a matter of debate whether disorder in $Pr_2Zr_2O_7$ induces transverse fields allowing quantum tunnelling among spin ice configurations[25], or is too strong and therefore pins most of the moment into the transverse components, the result being a paramagnetic state with quadrupolar correlations[24,26-27]. Recently, another QSI candidate, $Pr_2Hf_2O_7$, has emerged as a clean realization of a system of $Pr^{3+}$ Ising-like moments interacting on a perfect pyrochlore lattice[28]. We have produced large high-quality single crystals of $Pr_2Hf_2O_7$, by optimizing the conditions of the traveling solvent floating zone growth so that, despite the high melting temperature of about 2400 Celsius, the evaporation of praseodymium during this process is reduced as much as practically possible. As a result, the structural and magnetic properties are identical for our powder and single crystal samples, and there is no evidence for any sort of defects compared to a perfect cubic pyrochlore structure (see Supplementary Information).

In $Pr_2Hf_2O_7$, a sizeable crystal field gap, about 9 meV ~ 100 K, isolates a non-Kramers ground doublet with magnetic moment of ~ 2.4 $\mu_B$[28]. Temperatures much lower than the gap allow the description of the magnetic properties by an effective $S = 1/2$ local moment. The wave-function of this single-ion ground state doublet[28] leads to a strong Ising anisotropy at low temperature, as needed to stabilize a spin ice, and incorporates quadrupolar terms, which make it possible for the spin ice to acquire quantum dynamics through transverse exchange. Interestingly, the material displays no indication of symmetry-breaking order down to at least 0.05 K. However, a cooperative regime develops below 0.5 K, with macroscopic indications of spin ice correlations[28]. We present inelastic neutron scattering (INS) measurements, at approximately 0.05 K on our single crystals of Pr$_2$Hf$_2$O$_7$, in order to study the energy–momentum (*E*–***k***) dependence of the signals characterizing the correlated regime.

An overview of our data measured with unpolarized neutrons of incident energy $E_i$ = 2.7 meV is presented in Fig. 1, which shows constant-energy maps of the (H,H,L) plane in reciprocal space. Data on Fig. 1**a** were integrated around *E* = 0 ± 0.06 meV after subtraction of a background dataset measured at 50 K. The subtraction provides a good estimate of the magnetic part of the quasi-elastic structure factor at 0.05 K, which is directly proportional to the spin–spin correlation function. Bragg peaks that would indicate long-range magnetic order remain absent, but we observe a pattern of diffuse scattering that is characteristic of a spin liquid[1]. Our quasi-elastic structure factor of Pr$_2$Hf$_2$O$_7$ (Fig. 1**a**) has the general shape observed in pyrochlore materials with spin ice correlations, i.e. pinch points appear at Brillouin zone centres (0,0,2) and (1,1,1), extending into more diffuse scattering around wave-vectors ***k*** such as (0,0,3) and (3/2,3/2,3/2)[16]. The unambiguous presence of a quasi-elastic signal in our experiment contrasts with the observations made so far on samples of Pr$_2$Zr$_2$O$_7$, where the scattering appears mostly inelastic[23-25].

In Fig. 2 we plot the result of quasi-elastic cuts through the pinch point wave-vector (0,0,2), along the solid black lines shown in Fig. 1**a**, in order to diagnose the nature of the spin ice correlations in Pr$_2$Hf$_2$O$_7$. Our experiment indicates that pinch points appear in unpolarised scattering, which is expected in a near-neighbour classical spin ice (CSI) model (see the results of analytical calculations for a near-neighbour CSI – right part of Fig. 1**b** and red dashed lines in Fig. 2)[17]. The small magnetic moments in Pr$_2$Hf$_2$O$_7$ also suggest that it conforms to a near-neighbour CSI model rather than to a classical dipolar spin ice model where pinch points are only visible in the spin flip scattering of polarized neutrons (as in Ho$_2$Ti$_2$O$_7$, for example). However, pinch-point features in Pr$_2$Hf$_2$O$_7$ appear to be suppressed relative to the CSI model. Pinch-point scattering can be suppressed by quantum fluctuations[7], and therefore we test this hypothesis by comparing the data with the predictions of a lattice field theory of the photon excitations of a QSI[5]. The calculation of the equal-time QSI structure factor within the integration ranges of our experiment (left part of Fig. 1**b**) compares favourably with our data (Fig. 1**a**), with minor differences that we attribute to non-universal characteristics of Pr$_2$Hf$_2$O$_7$. Meanwhile, the two cuts shown in Fig. 2 reveal that both the experimental data (points with error bars), and the theoretical prediction for a QSI (blue solid line), show clear deviations from the predictions of a near-neighbour CSI. The best quantitative fits to the data are obtained using the QSI model for a temperature $T/ca_0^{-1} = 1.8 \pm 0.1$ ($a_0$ being the lattice constant of the pyrochlore cubic unit cell), which translates into a speed of light $c \approx 3.6 m/s$ for the artificial photons (assuming $T$ = 0.05 K). Such photon quasiparticles would appear in INS as gapless excitations with a bandwidth ∆E ≈ 0.01 meV, so that any photon spectral weight is integrated in our results of quasi-elastic scattering. This is consistent with the absence of low-lying dispersive excitations in the inelastic scattering (see below).

Fits of the quasi-elastic scattering favour the QSI model over the CSI model (Fig. 2), but are not by themselves decisive. Applying a standard "$\chi^2$" test, we find that it is statistically correct to choose the quantum theory over the classical one about 8 times out of 10 ($\chi^2$ is 55.8 for QSI, and 79.1 for CSI). This is encouraging, but to fully distinguish between the two models we should also look at other aspects of their physics, where the experimental signatures are expected to be drastically different. Here, inelastic scattering at higher energies is sharply discriminating. At the finite energy transfers in our experiments, the only excitations expected for a near-neighbour CSI are magnetic monopoles, which, if neutron-active, would appear as non-dispersing quasiparticle excitations with a unique energy. Meanwhile, it is well-established from theory that a continuum of scattering is expected from the quantum-coherent quasiparticles of a QSI.

The spectrum presented on Fig. 3**a** reveals a broad continuum of excitations present at 0.05 K. At the same time, the fact that spin flip scattering of neutrons polarized in the horizontal plane (X-SF) is a purely magnetic signal rules out the possibility that these excitations have a non-magnetic origin. Energy spectra taken in our INS data measured with unpolarised neutrons (Fig. 3**b**) confirm that this continuum of spin excitations extends up to at least $E$ = 1 meV. At low energy, the spectral weight is peaked around $E$ = 0.2 meV, consistent with our data taken on a powder sample at the same temperature[28]. When integrated around this value of energy transfer, the dynamical structure factor has a "starfish" form (Fig. 1**c**), but intensity exists at all wave-vectors. At energy transfers $E$ > 0.3 meV, the spectral intensity is broadly distributed throughout momentum space (see Supplementary Information, Fig. S3).

The existence of a gapped continuum of fractionalised excitations is one of the key signatures of emergent electrodynamics in a QSI[3-6]. Both the electric and magnetic charges of this theory are expected to act as propagating quasiparticles, with a dispersion set either by transverse exchange[29-31], or by the distribution of random transverse fields coming from disorder[32]; in either case, a broad continuum of excitations is expected in INS[33], but its form is not predicted in details that would allow a direct quantitative comparison with our experimental results. In $Pr_2Hf_2O_7$ we observe a continuum with a width approximately one order of magnitude higher than the effective exchange interaction, $J \approx 1.2$ K $\approx 0.1$ meV, which appears consistent with theoretical predictions[29-30,34]. The form of scattering found at finite energy (Fig. 1**c**) is highly reminiscent of quantum Monte Carlo simulations results for QSI at temperatures where there is a finite density of spinons[35]. It is also interesting to compare this behaviour with another candidate QSI, $Pr_2Zr_2O_7$. Features resembling pinch points have also been observed in INS data of $Pr_2Zr_2O_7$, with almost all spectral weight found above an energy $E \approx 0.2$ meV, and little or no elastic scattering[23-25]. Because of this, some authors have interpreted the scattering as evidence of a ground state with predominantly quadrupolar character[24,26-27]. $Pr_2Zr_2O_7$ also exhibits a broad continuum of excitations, extending up to an energy of 2 meV, the "lineshape" of which exhibits a scaling collapse, consistent with the splitting of the ground-state doublet of $Pr^{3+}$ by a broad distribution of random strain fields[25-26]. No such collapse is possible with INS data for $Pr_2Hf_2O_7$ powders[28] (see Supplementary Information), suggesting that the inelastic continuum cannot be explained in this way. A third possible explanation for the inelastic continuum in $Pr_2Hf_2O_7$ would be the existence of a hidden order, for example spin-nematic[36] or quadrupolar-ordered phases. However, since no thermodynamic phase transition is observed, this seems unlikely. By this process of elimination, which is set out in more detail in Section III of the Supplementary Information, we reach the conclusion that the continuum of scattering found in $Pr_2Hf_2O_7$ reflects the topological excitations of a QSI. In the absence of detailed predictions, the precise nature of these excitations remains an open question. Magnetic monopoles cannot usually be excited by scattering neutrons from a non-Kramers ion such as $Pr^{3+}$ [33], but could be introduced where disorder mixes dipolar and quadrupolar components of the ground state[32]. Alternatively, the continuum could originate in the dual, electric charges of the gauge theory[33].

Taken together, these results suggest that $Pr_2Hf_2O_7$ adopts a QSI ground state with fractionalized excitations – an important type of three-dimensional QSL. Our experimental observations display a striking resemblance to the predictions of a compact $U(1)$ gauge theory, exhibiting both the suppressed pinch points expected in the quantum coherent regime with emergent photons, and a continuum of excitations at higher energy. If confirmed, the identification of a QSI in $Pr_2Hf_2O_7$ is a significant breakthrough, and further experimental work is clearly warranted. Open challenges include measuring the $T^3$ contribution to heat capacity associated with linearly-dispersing photons[3,5]; determining the entropy associated with the spin-liquid state, a fundamental question in the context of the third law of thermodynamics[7]; determining how the suppression of pinch points evolves with temperature[5], which provides important information about quantum coherence in the QSI state; and directly probing the photons through higher resolution techniques, for example neutron spin echo. Although the disorder in our samples appears to be too weak to measure, it is still of interest to understand whether transverse exchange is the sole source of quantum fluctuations, or if disorder-induced transverse fields[32,25] also contribute to the quantum dynamics of $Pr_2Hf_2O_7$. On the theory side, predictions are needed concerning the structure factor of the spinons and a possible coexistence of the QSI with quadrupolar correlations. These developments would pave the way for exploring the possibilities of a genuine three-dimensional QSL.

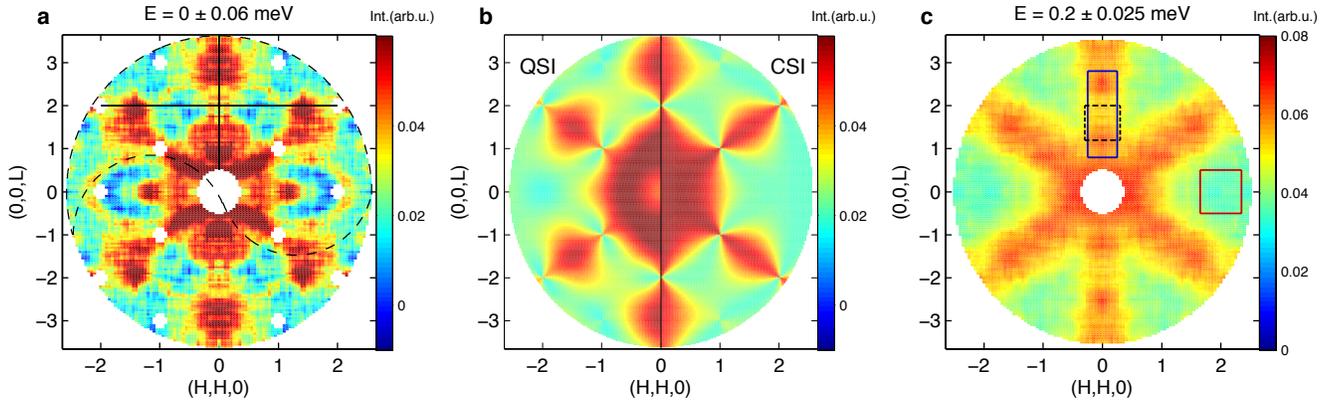

**Figure 1 | Momentum dependence of the magnetic correlations (a-b) and magnetic excitations (c) in $Pr_2Hf_2O_7$.** We show two-dimensional cuts in the (H,H,L) plane of reciprocal space, integrated around two different values of energy transfer. Data were collected using time-of-flight (ToF) inelastic neutron scattering (INS) at a temperature of 0.05 K on the IN5 instrument (unpolarised neutrons). The incident neutron energy was fixed to $E_i$ = 2.70 meV, providing an energy resolution at zero energy transfer of 0.05 meV (full width at half maximum). Data were measured in the upper quadrant shown by the black dashed curves on panel **a**, corrected for sample absorption and electronic noise, and then symmetrized (Methods). Maps shown on panels **a-c** are integrated over a thin range of momentum transfer, (H,-H,0) ± 0.05 r.l.u., in the direction perpendicular to the (H,H,L) plane. Panel **a** shows data that were integrated around zero-energy transfer (E = 0 ± 0.06 meV) after subtraction of a background dataset measured with the same statistics at a temperature of 50 K (all other conditions remaining unchanged). We refer to the data of panel **a** as 'quasi-elastic data', which, in our interpretation, mainly reflects the equal-time spin correlations of the quantum spin ice (QSI), including contributions from low-energy gapless photon excitations. The two continuous black lines in panel **a**, crossing each other at the wave-vector (0,0,2) where pinch point scattering is suppressed in a QSI, indicate the directions of the momentum dependent cuts shown on Fig. 2. Panel **b** shows the results calculated using a lattice field theory of a QSI (left part) and analytical calculations of a near-neighbour CSI (right part), for comparison with the experimental data of panel **a**. The QSI and CSI calculations are deduced from the fits shown on Fig. 2. Panel **c** shows inelastic data integrated around a finite value of energy transfer, E = 0.2 ± 0.025 meV. The background is negligible in the inelastic channels at E > 0.1 meV. The blue and red boxes on panel **c** are the integration areas in the data as they are shown here, of the energy spectra shown on Fig. 3**b**. The box delimited by black dashed lines on panel **c** of Fig. 1 represents the integration area that was used to generate the energy spectrum, shown on Fig. 3**a**, through polarized INS data (not shown).

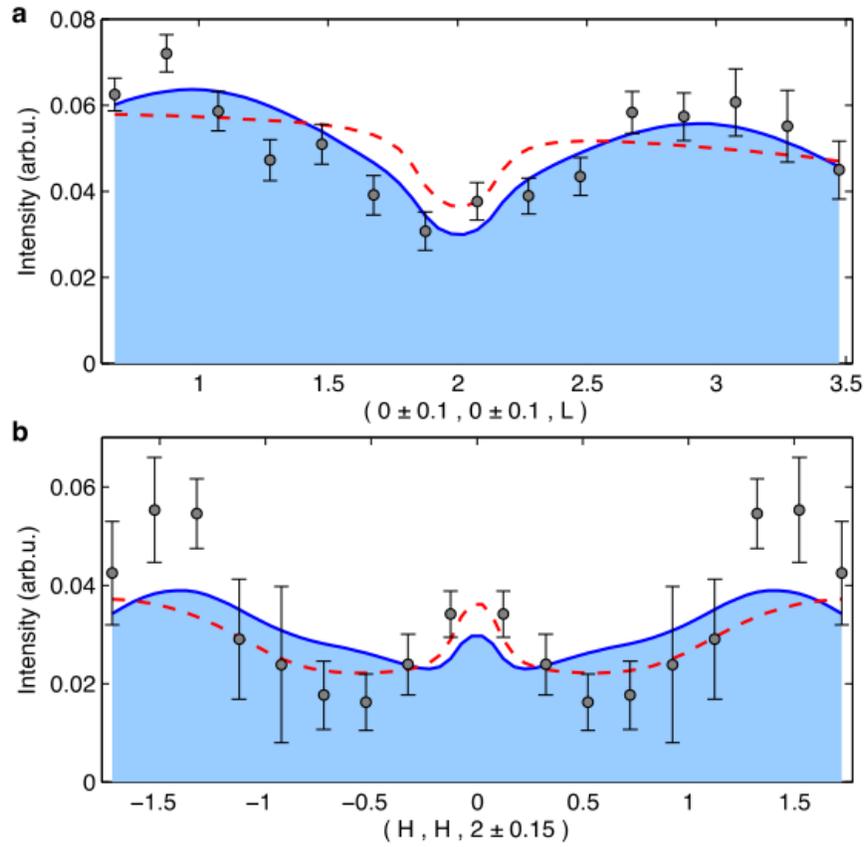

**Figure 2 | Line shape of the suppressed pinch points measured in Pr$_2$Hf$_2$O$_7$, and comparison with model calculations.** The data points with error bars show the results of radial (**a**) and transverse (**b**) cuts through the experimental data shown on Fig. 1**a**. The two cuts, whose directions are shown with black lines on Fig. 1**a**, cross each other at the zone centre (0,0,2), where a pinch point is expected to occur. The experimental data are compared with the prediction for both a classical near-neighbour classical spin ice (CSI, red dashed line) and a quantum spin ice[5] (QSI, solid blue line). Finite experimental resolution, modelled through an integration over a finite range of wave-vectors, leads to a small dip in the prediction for a classical spin ice around (0,0,2), visible in the transverse cut **a**, and eliminates a sharp spike at the same wave-vector in the transverse cut **(b)**. For the radial cut shown on panel **a**, quantum fluctuations lead to a further suppression of scattering around (0,0,2), as well as an enhancement of scattering around (0,0,1) and (0,0,3). For the transverse cut presented on panel **b**, quantum fluctuations lead to a non-monotonic evolution of scattering between H = 0.5 and H = 2. All of these QSI features are present in the experimental data, and the best fits for both cuts are obtained using a QSI model at a finite temperature $T/ca_0^{-1} = 1.8 \pm 0.1$, where $a_0$ is the lattice constant of the pyrochlore cubic unit cell and $c$ is the speed of light of emergent photons ($c \approx 3.6 m/s$ for T = 0.05 K). Further details of models and fits are provided in the "Methods" section. Error bars on panel **a** correspond to 1 standard error. For the symmetrized data points of panel **b**, we included the difference between two points before symmetrisation in order to reflect the fluctuation that caused the asymmetry, which is not counted in the statistical error for each individual point. The final error displayed on panel **b** is $E_{i(j)}^{sym} = (((I_i - I_j)/2)^2 + E_i^2 + E_j^2)^{1/2}$, where $I_{i,j}$ are the unsymmetrized intensities and $E_{i,j}$ are their individual statistical errors (1 standard error), respectively.

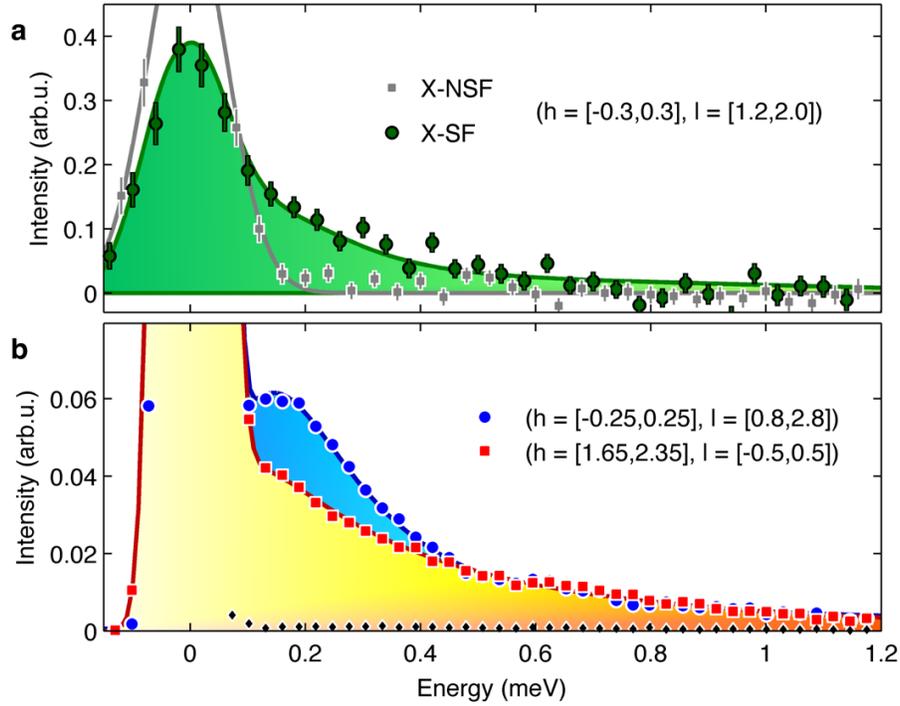

**Figure 3 | Energy spectra at fixed positions in momentum space.** We present constant-momentum cuts through our time-of-flight (ToF) inelastic neutron scattering (INS) data measured at a temperature of 0.05 K. The integration areas in momentum space are indicated with two vectors, h = [H,H,0] and l = [0,0,L], which correspond to the rectangles drawn on Fig. 1**c**. Data shown on panel **a** result from a polarized INS experiment realized on the instrument HYSPEC with an energy resolution of 0.125 meV at zero energy transfer (FWHM). We show the spin flip and non-spin flip scattering measured with neutrons that were polarized in the horizontal plane of the instrument, X-SF and X-SNF, respectively. The X-SF scattering is a purely magnetic signal in the absence of significant spin-incoherent scattering. This statement is justified in the case of our data because the inelastic NSF scattering, which contains 1/3 of the total spin-incoherent signal, is negligible, and, therefore, the 2/3 of the spin-incoherent signal expected in the SF scattering cannot give a sizeable contribution to our magnetic SF signal. The elastic NSF response is likely from nuclear isotope incoherent scattering. The data on panel **a** demonstrate the existence of quasi-elastic and inelastic (over the entire range of accessible energy transfers $E$) signals that are, unambiguously, magnetic scattering. On panel **b** we show the energy spectra through the unpolarised INS data measured on IN5 (data shown on Fig. 1) with an energy resolution of 0.050 meV at zero energy transfer (FWHM). The inelastic scattering peaks at $E$ = 0.2 meV, and scattering at these energies is highly structured, showing a "starfish" pattern in the [H,H,L] plane (Fig. 1**c**). Above $E$ = 0.3 meV, the inelastic signal becomes too weak to accurately resolve its wave-vector dependence (see Supplementary Information, Fig. S3). The integration in two specific areas of reciprocal space, at and away from the "starfish" pattern (blue and red symbols, respectively), provides evidence for a continuum of excitations, which we attribute to the fractionalized excitations of a QSI ground state. The black symbols on panel **b** show an energy spectrum through data collected at a temperature of 50 K, scaled by the ratio of the Bose factors at 50 K and 0.05 K, which gives an estimate of the inelastic background at 0.05 K. All lines are guides to the eye. Error bars correspond to 1 standard error.

**Acknowledgements**

We acknowledge the Institut Laue Langevin, ILL (Grenoble, France, EU) for the allocated beamtime. We acknowledge funding from the Swiss National Science Foundation (Grants No. 200021_140862; No. 206021_139082 and No. 200021_138018). This research used resources at the Spallation Neutron Source, a DOE Office of Science User Facility operated by the Oak Ridge National Laboratory. The work at ORNL was supported by the U.S. Department of Energy, Office of Science, Office of Basic Energy Sciences, under contract number DE-AC05-00OR22725. The work at the University of Warwick was supported by the EPSRC, UK, through Grant EP/M028771/1. Additional neutron scattering experiments were carried out at the continuous spallation neutron source SINQ at the Paul Scherrer Institut at Villigen PSI in Switzerland.

**Author contributions**

Project and experiments were designed by R.S., T.F. and M.K. Crystal growth and characterization were performed by R.S., M.C.H. and G.B. Sample alignment and mounting for the neutron scattering experiment was realized by R.S. and N.G. Neutron scattering experiments were carried out by R.S. and N.G. with J.O. and B.W. as local contacts. The experimental data were analysed by N.G., R.S., T.F. and M.K. Calculations were made by H.Y. and N.S. The paper was written by R.S. with feedback from all authors.

**Competing financial interests**

The authors declare no competing financial interests.

**Methods**

**Sample preparation**

A large single-crystal of $Pr_2Hf_2O_7$ was grown by the floating zone technique using an optical furnace equipped with high-power xenon-arc lamps[37]. The sample was characterized by synchrotron X-ray powder diffraction for a precise determination of the lattice parameter[37]. Additional single-crystal neutron diffraction data were collected on the Zebra instrument, a technique that provides an excellent contrast between the three atoms present in $Pr_2Hf_2O_7$. Attempts to refine antisite cation disorder and oxygen Frenkel disorder, which can induce stuffing effects and disordered interactions and environments, respectively, did not provide evidence for structural defects.

**Neutron scattering experiments**

The unpolarised inelastic neutron scattering experiment was performed with the IN5 time-of-flight spectrometer at the Institut Laue-Langevin, Grenoble, France[38]. The single crystal, mounted on a copper sample holder and fixed with copper wires, was aligned in the (H,H,L) plane in reciprocal space. Measurements were carried out at a temperature T = 0.05 K with a fixed incident neutron energy $E_i$ = 2.70 meV and an energy resolution of 0.050 meV at zero energy transfer (FWHM). The measurements were taken while rotating the sample about the vertical axis. The data analysis was performed using HORACE analysis software[39]. The raw data were corrected for time-independent electronic noise, and for neutron absorption by a finite element analysis based on the sample geometry. For the quasi-elastic scattering results, measurements at T = 50 K were used for background subtraction. The presented data are symmetrized: they were folded in the first quadrant to improve statistics, unfolded to cover the four quadrants and smoothed.

The polarized inelastic neutron scattering experiment was performed with the HYSPEC time-of-flight spectrometer[40] at the Spallation Neutron Source, Oak Ridge National Laboratory, USA. The same single crystal measured with the IN5 spectrometer was oriented in the (H,H,L) plane in reciprocal space on a copper sample holder and fixed with copper wires. Measurements were carried out at a temperature T = 0.05 K with an incident neutron energy $E_i$ = 3.8 meV and an energy resolution of 0.125 meV at zero energy transfer (FWHM). The neutrons were polarized using a Heusler crystal array, and analysed with the polarization analyser constituted of a supermirror array designed and built at the Paul Scherrer Institut[40]. A Mezei flipper located between the monochromator and the sample was used to flip the polarization and measure both spin flip and non-spin flip scattering. A multi-coil electromagnet surrounding the sample was used to rotate the incident vertical neutron polarization to a polarization in the horizontal scattering plane. In the standard notation of neutron polarization analysis, the polarization X describes a neutron spin polarized parallel to the momentum transfer **k**, and in these conditions all magnetic scattering events occur through spin-flip scattering[41]. In the reported experiment, the polarization direction in the horizontal plane was fixed to be parallel to the momentum transfer **k**, when it takes a value |**k**| = 1.02 Å$^{-1}$. The data reported here are integrated in a small region in reciprocal space around |**k**| = 1.02 Å$^{-1}$ and therefore represent X-polarized scattering in the standard notation of neutron polarization analysis. Measurements of the sample holder without the sample were used for background subtraction. Data were corrected for the flipping ratio of the neutron spin with measurements of spin flip and non-spin flip scattering of quartz. The mirror-dependent transmission of the supermirror array was calibrated by using vanadium as a standard. The energy-dependent mirror transmission was also corrected based on previous calibration of the supermirror array.

**Fitting of the experimental data to model calculations**

The analytical predictions of $S(\boldsymbol{q})$ from Quantum Spin Ice (QSI) in Fig. 1 and Fig. 2 are calculated with the $U(1)$ lattice-gauge theory described in Ref. 5. The $S(\boldsymbol{q})$ from Classical Spin Ice (CSI) model are calculated based on Refs. 5, 42 and 43. The QSI model corresponds to the limit $U \gg W$ in Ref. 5, while the CSI model corresponds to the limit $W \gg U$.

In both Fig. 1 and Fig. 2, the magnetic form factor of Pr$^{3+}$ in the dipolar approximation is included to be consistent with experiments. Furthermore, for Fig. 2, the analytical results are integrated over the same range as the experimental data, which is $(\pm 0.1, \pm 0.15)$ for $(H, L)$ inside $(H, H, L)$ plane, and $\pm 0.05$ perpendicular to it.

In order to fit the experimental data with the QSI model in Fig. 2, we used the speed of artificial light $c$ as a fitting parameter, which depends on microscopic details of the material, and overall rescaling and shift of the intensity due to the arbitrary unit of the intensity and paramagnetic background subtraction of experiments, respectively. The optimal value of $c$, overall rescaling, and shift were obtained through a weighted least-square fit of $\sum_{i \in \text{all data points}} \left(I_i^{\text{exp}} - I_i^{\text{thy}}\right)^2 / E_i^2$, where $I_i^{\text{exp}}$ is the experimental intensity, $I_i^{\text{thy}}$ is the theoretical prediction, and $E_i$ the error associated with each data point. For the symmetrized data points of Fig. 2**b**, we included the difference between two points before symmetrisation in order to reflect the fluctuation that caused the asymmetry, which is not counted in the statistical error for each individual point. The final error is $E_{i(j)}^{\text{sym}} = \left(\left((I_i - I_j)/2\right)^2 + E_i^2 + E_j^2\right)^{1/2}$, where $I_{i,j}$ are the unsymmetrized intensities, and $E_{i,j}$ are their individual statistical errors respectively.

The CSI model is fitted in the same way as the QSI model, but with only the overall rescaling and shift as adjustable parameters. The zero-energy slice of neutron scattering for the CSI model corresponds to our analytical calculation of zero temperature, leaving no further adjustable parameters.

The speed of emergent photon obtained by the fitting is $c = 3.6 \, m/s$, and bandwidth $\Delta E \approx 0.01 \, meV$, meaning that the energy-integrated $S(\boldsymbol{q})$ of the QSI is within experimental resolution and is the proper quantity to use for the fit. We note that the $U(1)$ lattice-gauge theory described in Ref. 5 addresses the emergent photons of a QSI, and makes explicit predictions for the dynamical structure factor $S(\boldsymbol{q}, E)$ measured in experiments. The corresponding equal-time structure factor $S(\boldsymbol{q})$ is found by integrating this dynamical structure over the photon bandwidth $\Delta E \approx 0.01 \, meV$. Therefore, the signal measured in the elastic channels of our IN5 experiment should properly be regarded as a 'quasi-elastic' signal integrating over $\Delta E \approx 0.01 \, meV$.

The weighted square sum is 55.8 for QSI and 79.1 for CSI, showing significant improvement of fitting when quantum effects are switched on. The dip in the centre of classical prediction shown in Fig. 2**a** is purely a consequence of the integration over experimental resolution, and its overall slope is from the magnetic form factor. Without these effects, the classical prediction would be a constant, independent of the wave-vector, on this cut through reciprocal space, while the quantum theory shows suppressed scattering near the centre (0,0,2) and enhanced scattering near (0,0,1) and (0,0,3).

The speed of light $c$, overall rescaling, and shift obtained through least square fitting of the QSI to the experimental data are then used to produce the analytical results of Figure 1**b**.

**Data availability**

The data that support the plots within this paper and other findings of this study are available from the corresponding author upon reasonable request. The datasets for the inelastic neutron scattering experiment on IN5 are available from the Institute Laue-Langevin data portal (doi: 10.5291/ILL-DATA.4-05-641).[44]

**References (methods)**